\documentclass[aps,prb,amsmath,superscriptaddress,twocolumn,longbibliography]{revtex4-1}

\usepackage{graphicx}
\usepackage{amsmath}
\usepackage{amsfonts}
\usepackage{amssymb}
\usepackage{braket}
\usepackage{bm}
\usepackage{multirow}
\usepackage{color}
\usepackage[normalem]{ulem}
\usepackage{mathrsfs}
\usepackage{mathtools}
\usepackage{float}
\usepackage{enumerate}

\newcommand{\blue}[1]{{\textcolor{blue}{#1}}}

\makeatletter

\newcommand{\Rmnum}[1]{\expandafter\@slowromancap\romannumeral #1@}
\makeatother

\usepackage[breaklinks]{hyperref}
\hypersetup{colorlinks=true, linkcolor=blue, citecolor=blue, filecolor=blue, urlcolor=blue}

\AtBeginDocument{%
    \newwrite\bibnotes
    \def\bibnotesext{Notes.bib}
    \immediate\openout\bibnotes=\jobname\bibnotesext
    \immediate\write\bibnotes{@CONTROL{REVTEX41Control}}
    \immediate\write\bibnotes{@CONTROL{%
    apsrev41Control,author="08",editor="1",pages="1",title="0",year="1"}}
     \if@filesw
     \immediate\write\@auxout{\string\citation{apsrev41Control}}%
    \fi
}%

\def\sz{\sigma^{\rm z}}
\def\sx{\sigma^{\rm x}}

\def\mz{\mu^{\rm z}}
\def\mx{\mu^{\rm x}}

\def \nz{\nu^{\rm z}}
\def \nx{\nu^{\rm x}}

\def\ket#1{|\, #1\,\rangle}

\def\openone{{1\!\!1}}

\def\v#1{v(#1)}
\def\vtil#1{{\tilde v}(#1)}

\def\Q8{{Q}_8}

\begin{document}

\title{Constructing Non-Abelian Quantum Spin Liquids\\ Using
  Combinatorial Gauge Symmetry}

\author{Dmitry Green}
\email{dmitry.green@aya.yale.edu}
\affiliation{Physics Department, Boston University, Boston, MA, 02215, USA}
\affiliation{AppliedTQC.com, ResearchPULSE LLC, New York, NY 10065, USA}

\author{Claudio Chamon}
\email{chamon@bu.edu}
\affiliation{Physics Department, Boston University, Boston, MA, 02215, USA}

\date{\today}

\begin{abstract}
  We construct Hamiltonians with only 1- and 2-body interactions that
  exhibit an {\it exact} non-Abelian gauge symmetry (specifically,
  combinatiorial gauge symmetry). Our spin Hamiltonian realizes the
  quantum double associated to the group of quaternions. It contains
  only ferromagnetic and anti-ferromagnetic $ZZ$ interactions, plus
  longitudinal and transverse fields, and therefore is an explicit
  example of a spin Hamiltonian with no sign problem that realizes a
  non-Abelian topological phase. In addition to the spin model, we
  propose a superconducting quantum circuit version with the same
  symmetry.
 \end{abstract}

\maketitle

\section{Introduction}
\label{sec:intro}

Non-Abelian topological states are some of the most remarkable forms
of quantum matter. The exchange of quasiparticle excitations in these
systems is characterized by non-commuting unitary transformations in a
space of degenerate many-body states, i.e., these quasiparticles have
non-Abelian braiding statistics~\cite{Frolich, Witten}. Non-Abelian
states are theoretically predicted to describe certain fractional
quantum Hall (FQH)
states~\cite{Wen,MooreRead,ReadRezayi,HaldaneRezayi}. Kitaev's
honeycomb spin-liquid model~\cite{hex_Kitaev} is another example; it
displays a non-Abelian phase in the presence of a magnetic field, with
excitations that have Ising-anyon statistics. A more general class of
systems that realize non-Abelian topological states of matter is that of
Kitaev's exactly solvable quantum double models~\cite{Kitaev}, in
which the specific state is determined by the choice of the
non-Abelian group in which the link (or gauge) degrees of freedom take
their values.

An obstacle to realize the quantum double models in experimental systems is that they are written in terms multi-body interactions among degrees of freedom expressed as group elements, not physical degrees of freedom, such as spins or charges. To implement the quantum doubles experimentally would require designing parent Hamiltonians with 1- and 2-body interactions. Notable efforts
  along these lines have been made in
  Refs.~\onlinecite{Ioffe_NonAbelian,Ludwig}
  and~\onlinecite{Yoshida}. The local gauge symmetries in quantum
  double realizations of Refs.~\onlinecite{Ioffe_NonAbelian,Ludwig}
  are emergent, being active only in the low energy sector of the
  theory (hence perturbative). On the other hand, the local gauge
  symmetries are exact in the case of Ref.~\onlinecite{Yoshida}, but
  it is not clear that the Hamiltonian is physically realizable like
  in Ref.~\onlinecite{Ioffe_NonAbelian}, where a physical
  implementation is proposed using arrays of Josephson junctions. The
  goal of this paper is to develop a framework that fills in the gaps in both of these approaches: we
  design a physical Hamiltonian with exact local non-Abelian gauge
  symmetries, using only 1- and 2-body interactions that could be
  implemented in physical systems, such as superconducting quantum
  circuits.

The program hinges on extending combinatorial gauge
symmetry~\cite{Z2-CGS} (see Ref.~\onlinecite{Abelian-CGS} for an
  in depth introduction to the symmetry principle for Abelian theories
  that is accompanied by step-by-step constructions of examples) to a
non-Abelian theory. The gauge symmetries are built into the {\it
  microscopic} Hamiltonian, and hence are {\it exact}, as opposed to
emerging only in a low energy limit. That the gauge symmetry is exact
in realistic Hamiltonians expands the range of parameters for which
the topological phase may be stable, thus providing a way to escape
limits on the sizes of the attainable energy gaps. Moreover, the model
has ferromagnetic and anti-ferromagnetic $ZZ$ interactions, plus
longitudinal and transverse fields. Therefore, the spin model is an
{\it explicit} realization of a spin Hamiltonian without a sign
problem that realizes a non-Abelian topological phase.

We focus on the quantum double with link (or leg) variables taking
values within the quaternion group, $\Q8$, on a honeycomb lattice. We
represent the 8 quaternion variables ($\pm 1, \pm i, \pm j$, and
$\pm k$) with spin-1/2 degrees of freedom. We shall utilize 4
``gauge'' spins in each link of the honeycomb lattice, thereby
defining a 16-dimensional Hilbert space that we split into two sets,
of even and odd parity states, and use the 8 even parity states to
represent the 8 quaternions. The construction utilizes ``matter''
spins on the links to split the even and odd parity states and on the
sites to enforce that the three quaternion variables multiply to the
identity (``zero flux'' condition).

Finally, we present a superconducting quantum circuit with the same
non-Abelian combinatorial gauge symmetry. In the limit where the
superconducting wires are small, and voltage biases are tuned so that
two nearly degenerate charge states are favored in each wire, the
system becomes a non-Abelian generalization of the WXY model
introduced in Ref.~\onlinecite{sc-waffle}. In this case, the remaining
energy scale in the problem is the Josephson coupling, and if the
system (with the combinatorial gauge symmetry) is gapped, the
non-perturbative gap must be necessarily on the order of this scale.

\section{Preliminaries and a roadmap}
\label{sec:roadmap}

Before diving into details, we think that it will aid the presentation
to very briefly summarize the basic elements of the quantum
double~\cite{Kitaev} and how we will use them. For an even more
general introduction to discrete gauge theories see
Ref.~\onlinecite{PropitiusBais} (this reference also includes an
explicit construction of excitations and their fusion rules for the
quaternion group).

\begin{figure*}[th]
\centering
\includegraphics[width=0.7\textwidth]{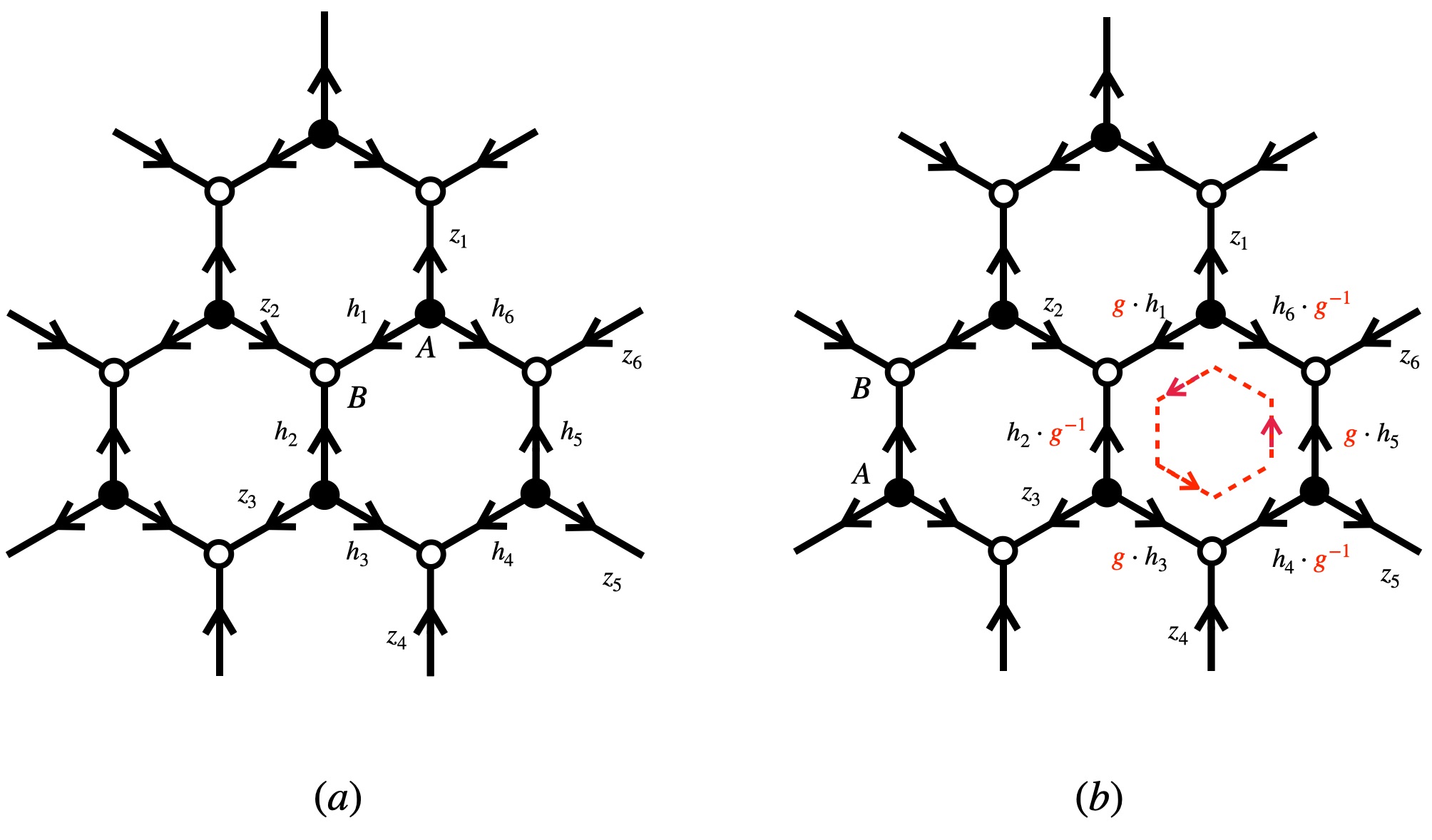}
\caption{Representation of a finite non-Abelian gauge theory on the
  hexagonal lattice. (a) Each link is oriented and a degree of
  freedom, taking values on a group $G$, lives on it. Examples of link
  degrees of freedom are those labeled by $z_i\in G$ and $h_i\in
  G$. We have chosen link orientations such that sublattices $A$ and
  $B$ have all arrows pointing out and in, respectively. In the ground
  state the zero ``flux'' on each vertex is equivalent to either
  clockwise (sublattice $A$) or counterclockwise (sublattice $B$)
  multiplication of the group elements on the associated legs to the
  identity. (In the ground state, $z_1\cdot h_6\cdot h_1=1$ and
  $z_2\cdot h_2\cdot h_1=1$ on $A$ and $B$, respectively.) The order
  of multiplication is important as the group is non-Abelian. (b)
  Local gauge symmetry associated with a group element $g$ is shown in
  red. The plaquette is denoted by an oriented path in red. Each time
  a link is traversed in the same direction as its orientation the
  group element on the leg is multiplied by $g$ from the
  left. Conversely, traversing along a leg against its orientation
  multiplies the group element on the leg by $g^{-1}$ from the
  right. The flux is preserved on each vertex around the
  plaquette. Note again that the order matters because the group is
  non-Abelian.}
\label{fig:Lattice} 
\end{figure*}

The Hilbert space for a finite gauge theory on a lattice is spanned by
states $|z\rangle$ on each link, where $z$ is an element in a group
$G$. Each link also has an (arbitrary) orientation which, for a
non-Abelian group, is necessary to define how operators act within the
Hilbert space. See Fig.~\ref{fig:Lattice}.

Conventionally, the notion of flux is defined by a product of the group
elements around plaquettes. In our approach we find it convenient to
use the dual lattice, instead, where flux is defined on the vertices
by the product of group elements on the legs around the vertex. We
refer the reader to Kitaev's work on the quantum double for details.

The gauge symmetry can be thought of as the insertion of any group
element $g$ into any plaquette. In this operation the states on the
links surrounding the plaquette are multiplied in a specific order
such that the flux at each vertex is preserved. Since the group
elements do not commute, some link states are multiplied from the left
and some from the right (see Fig.~\ref{fig:Lattice}). The Hamiltonian
is invariant under any such plaquette transformation.

The core challenge in realizing such a model is two-fold. First, what are
the physical elements that can be used to represent the abstract group
elements $g$? Second, how do we construct a Hamiltonian that is
invariant under these abstract operations where vertex and plaquette
terms are products of multi-leg terms?

We are able to address both challenges for the quaternion group on a
hexagonal lattice. For this case, we find a representation of both the
group elements and the Hamiltonian by ordinary spin-1/2 states that
are coupled by simple Ising interactions.

At the heart of our approach is combinatorial gauge
symmetry. Schematically, suppose we have an Ising Hamiltonian that
couples two sets of spins $\mu^z_n$ and $\sigma^z_i$. The $\sigma$ are
what we call gauge spins and $\mu$ are matter spins. All the gauge
symmetries that we want to emulate are embodied by the gauge spins,
while the matter spins serve to enable the symmetry via permutations of states within the enlarged Hilbert space.

\begin{figure}[!ht]
\centering
\includegraphics[width=0.4\textwidth]{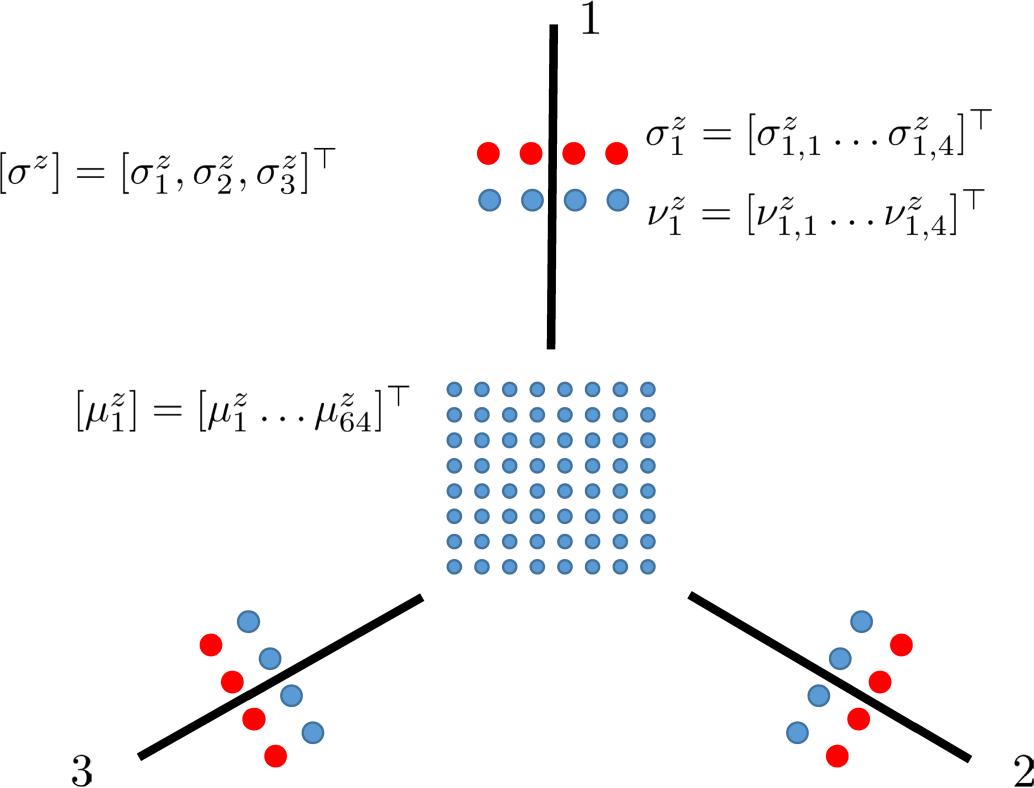}
\vspace{0.5cm}
\caption{Spin construction for the system with non-Abelian gauge
  symmetry. Spin-1/2 degrees of freedom are placed on links and on the
  vertices of the lattice (one vertex and three links or legs are
  shown). Gauge spins are depicted in red, and matter spins, placed
  both at the links and at the vertex, are depicted in blue. There are
  two steps in the construction. First, the Hilbert space of
  quaternions on each leg $\alpha=1,\ldots,3$ is implemented by a set
  of gauge spins, $\sz_\alpha$, and a companion set of four matter
  spins, $\nz_\alpha$, that are coupled. Each of $\sz_\alpha$ and
  $\nz_\alpha$ are 4-spinors and the specific form of the coupling
  between them serves to implement the quaternion group at each
  site. The combined 12-spinor of gauge spins is denoted by
  $[\sz]$. Second, all $12$ gauge spins are coupled to an additional
  set of $64$ matter spins $\mu^{\text z}_n$ ($n=1,\ldots,64$) that
  are at the center of the vertex. The coupling is all-to-all, i.e.,
  there are $12\times 64$ couplings between $[\sz]$ and $[\mz]$. The
  specific form of this coupling will serve to implement the local
  quaternion gauge symmetry.  Note that the matter spins $\nz$ and
  $\mz$ are not coupled and that the gauge spins are shared by
  neighboring vertices in the sense that they couple to matter spins
  $\mz$ residing at the center of neighboring vertices.}
\label{fig:geometry} 
\end{figure}

Consider the general Hamiltonian at a given site $s$ (details do not
matter for the purposes of the roadmap):
$H_s=\sum_{ni}\mu^z_n\; W_{ni}\; \sigma^z_i$ with $n=1,\ldots,p$ and
$i=1,\ldots,q$. Now suppose we want to represent a given group
operator $g$. We construct a $q\times q$ representation matrix $R(g)$
that acts on the set of $\sigma^z_i$ two legs at a time. For a
suitably chosen set of interactions $W_{ni}$ we require that there is
a companion $p\times p$ permutation matrix $L(g)$ that acts on the
matter spins $\mu^z_n$ such that the Hamiltonian is invariant. In
other words we demand the automorphism $L^{-1}(g)\; W\; R(g) = W$ for
all $g$ on all sites $s$ and hence gauge invariance. A key feature is
that both $L$ and $R$ transformations must be monomial matrices in
order to preserve the spin commutation relations. In this way we are
able to represent all operations of the group elements $g$ on the
gauge spins by absorbing them into permutations of matter spins. This
symmetry is exact by construction. In effect we have reduced the
problem to that of finding an appropriate $W$ and representations of
$R$ and $L$ in terms of spins.

Once we have constructed the appropriate representation we will need
to ensure that the ground state manifold is in fact composed only of
the states that satisfy the ``zero flux'' condition that the group
elements on a star multiply to the identity. We will find that it is
necessary to add a longitudinal field on some of the matter spins
(respecting the combinatorial symmetry), but this is a straightforward
interaction from a physical point of view.

The geometry of our solution is shown in Fig.~\ref{fig:geometry}; we
describe it in detail below.

\section{Spin representation of quaternions and left/right representations}
\label{sec:symmetry}

\begin{figure}[bh]
\centering
\includegraphics[width=0.45\textwidth]{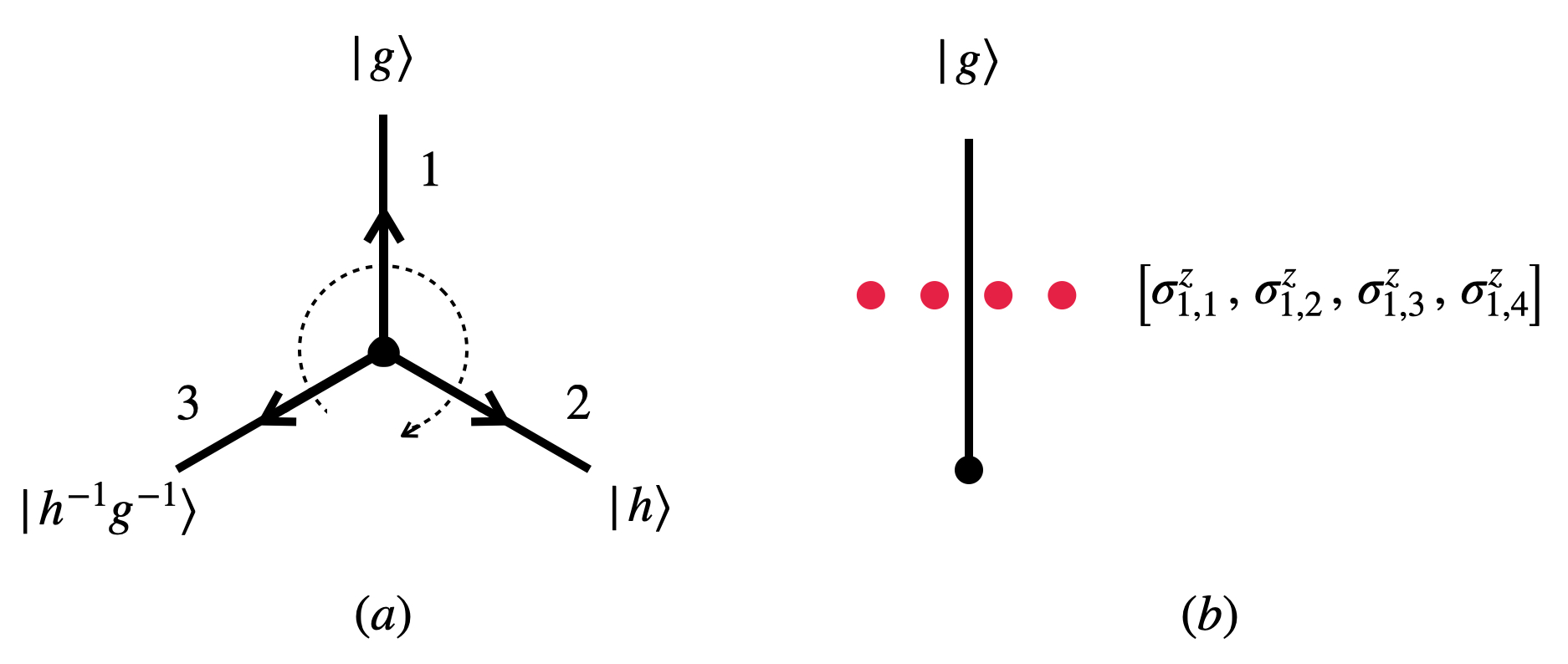}
\caption{(a) Dual lattice version of Kitaev's quantum double. The
  ground states are defined by configurations in which the product of
  the allowed group elements clockwise (orientation matters) is
  $g_1g_2g_3=1$, where $g_1=g$, $g_2=h$ and
  $g_3=(gh)^{-1}=h^{-1}g^{-1}$. (b) The quaternion group element on
  each leg is represented by a 4-vector of the eigenvalues $(\pm)$ of
  four spins. The figure shows leg 1 as an example, with the red dots
  representing the four gauge spins (matter spins are not shown for
  simplicity).}
\label{fig:QuantumDouble} 
\end{figure}

Consider quaternion degrees of freedom placed on the links of a
honeycomb lattice. Let us focus on the vertices of the lattice, as
depicted in Fig.~\ref{fig:QuantumDouble}(a), and construct interactions
that favor the configurations in which the three link variables
$g_1,g_2,g_3\in \Q8$ multiply (clockwise) to the group identity:
$g_1\,g_2\,g_3=1$. (Note that we focus on the 3-legged stars
of the honeycomb lattice, as opposed of the 3-sided plaquettes of a
triangular lattice as in Kitaev's original formulation; these two
systems are equivalent, just formulated in dual lattices.) We
represent the quaternion elements using 4 spin-1/2 variables, as
illustrated in Fig.~\ref{fig:QuantumDouble}(b).

The Hilbert space for the link variable is spanned by the states
$\ket{g}$, with $g\in \Q8$. The formulation of the quantum double of
Ref.~\onlinecite{Kitaev} defines operators
$L_+^h\ket{g}\equiv \ket{hg}$ and $L_-^h\ket{g}\equiv \ket{gh^{-1}}$
that multiply the group element inside the ket on its left or
right. In representing the elements of the quaternion group in terms
of spins, we shall define monomial matrices $\ell(h)$ and $r(h)$ that
will have similar effects on the spin states (see
Ref.~\onlinecite{Moore} for an exercise on writing left/right
representations of $\Q8$; we introduce and use different matrices here that are suitable for basis vectors that contain only $\pm 1$ elements that represent spins rather than $0$'s and $1$'s.)

Let us associate the following 4-vectors to the group elements of $\Q8$:
\begin{alignat}{3}
  \v{+1}&=
  \begin{bmatrix}
    ++++
  \end{bmatrix}
  \qquad
  \v{-1}&=
  \begin{bmatrix}
    ----
  \end{bmatrix}
  \nonumber\\
  \v{+i}&=
  \begin{bmatrix}
    +-+-
  \end{bmatrix}
  \qquad
  \v{-i}&=
  \begin{bmatrix}
    -+-+
  \end{bmatrix}
  \nonumber\\
  \v{+j}&=
  \begin{bmatrix}
    ++--
  \end{bmatrix}
  \qquad
  \v{-j}&=
  \begin{bmatrix}
    --++
  \end{bmatrix}
  \nonumber\\
  \v{+k}&=
  \begin{bmatrix}
    -++-
  \end{bmatrix}
  \qquad
  \v{-k}&=
  \begin{bmatrix}
    +--+
  \end{bmatrix}
  &\;.
  \label{eq:vector_rep}
\end{alignat}
The $\pm$ stand for $\pm1$. We will use spins to represent each entry
in the vectors $v$, in which case the $\pm$ should be thought of as
representations of the two eigenstates in the $\sz$ basis:
$+\equiv(1\;0)$ and $-\equiv(0\;1)$. Notice that all $v$ have even
parity.

One can represent the action of left and right multiplication by group
elements through matrices $\ell(g)$ and $r(g)$ such that
\begin{align}
  v(g) \;\ell(h) &= v(hg)
  \nonumber
  \\
  v(g) \;r(h) &= v(gh)
  \;.
\label{eq:rl_multiplication}
\end{align}

\begin{widetext}
The $4\times 4$ monomial matrix representations for these left/right operators
are
\begin{subequations}
\label{eq:l-and-r-4x4}
\begin{align}
  r({\pm 1})=
  \begin{bmatrix}
    \pm&0&0&0\\
    0&\pm&0&0\\
    0&0&\pm&0\\
    0&0&0&\pm
  \end{bmatrix}
  \quad
  r({\pm i})=
  \begin{bmatrix}
    0&\mp&0&0\\
    \pm&0&0&0\\
    0&0&0&\mp\\
    0&0&\pm&0
  \end{bmatrix}
  \quad
  r({\pm j})=
  \begin{bmatrix}
    0&0&0&\mp\\
    0&0&\mp&0\\
    0&\pm&0&0\\
    \pm&0&0&0
  \end{bmatrix}
  \quad
  r({\pm k})=
  \begin{bmatrix}
    0&0&\pm&0\\
    0&0&0&\mp\\
    \mp&0&0&0\\
    0&\pm&0&0
  \end{bmatrix}
             \;{}
\end{align}
and
\begin{align}
  \;\;\ell({\pm 1})=
  \begin{bmatrix}
    \pm&0&0&0\\
    0&\pm&0&0\\
    0&0&\pm&0\\
    0&0&0&\pm
  \end{bmatrix}
  \quad
  \ell({\pm i})=
  \begin{bmatrix}
    0&0&0&\mp\\
    0&0&\pm&0\\
    0&\mp&0&0\\
    \pm&0&0&0
  \end{bmatrix}
  \quad
  \ell({\pm j})=
  \begin{bmatrix}
    0&0&\mp&0\\
    0&0&0&\mp\\
    \pm&0&0&0\\
    0&\pm&0&0
  \end{bmatrix}
  \quad
  \ell({\pm k})=
  \begin{bmatrix}
    0&\pm&0&0\\
    \mp&0&0&0\\
    0&0&0&\mp\\
    0&0&\pm&0
  \end{bmatrix}
             \;.
\end{align}
\end{subequations}
\end{widetext}
Note that both $r$ and $\ell$ preserve the parity of
each $v$ because they are monomial matrices with an even number of
$-1$'s. Also note that $\ell(h^{-1})=\ell^\top(h)$ and
$r(h^{-1})=r^\top(h)$. In terms of their action on the spin degrees of
freedom, the $+$ and $-$ signs inside these matrices should be
interpreted as $\openone_2$ and $\sx$ operators acting on the
underlying spins; the latter flips a spin and the former is the
$2\times 2$ identity operator.

One can also construct a convenient $4\times 4$ matrix
$w$ that implements inversion of a group element $h$:
\begin{align}
  v(h)\; w = v(h^{-1})
  \;.
\end{align}
$w$ turns out to be the Hadamard matrix
\begin{align}
  w
  =\frac{1}{2}\;
  \begin{bmatrix}
    -&+&+&+\\
    +&-&+&+\\
    +&+&-&+\\
    +&+&+&-
  \end{bmatrix}
  \;.
  \label{eq:leg_w}
\end{align}
This matrix ties the left/right representations together:
\begin{align}
    w \;r(h)\; w = \ell(h^{-1})
    \quad\text{and}\quad
    w \;\ell(h)\;w = r(h^{-1})\;.
\label{eq:gauge}
\end{align}
Crucially, it follows that $w$ is invariant under left/right monomial
transformations,
\begin{align}
  \ell(h)\;w \;r(h)  = w
  \quad\text{and}\quad
  r(h)\;w \;\ell(h) = w
  \;.
\label{eq:w-invariance}
\end{align}
This is a key relation, and it is one of two automorphisms that we will use in our construction of the Hamiltonian (per Sec.~\ref{sec:roadmap}).

\section{Building the quaternion quantum double Hamiltonian}

We construct the Hamiltonian for the quaternion quantum double in two
steps. The first step is to construct terms to separate 16 states
associated with 4 spin-1/2 degrees of freedom, in each link, into two
sets of 8 states each. One set of states -- those with even parity --
will correspond to the representation of the elements of $\Q8$ as
described above; the other set of 8 states -- those with odd parity --
will be pushed up in energy\blue{, as discussed in Sec.~\ref{sec:links}
below}.

The second step of the construction is to design \blue{in
  Sec.~\ref{sec:vertex}} a Hamiltonian that is\blue{, as shown in
  Sec.~\ref{sec:sym},} invariant under transformations associated to
left/right multiplication of the link variables on a vertex by group
elements that leave vertices where the product $g_1\,g_2\;g_3=1$
invariant, as illustrated in Fig.~\ref{fig:VertexTransformation}. We
note that the quaternion combinatorial gauge symmetry itself, as we
shall see, is independent of the even-odd splitting, so the
construction is non-perturbative, as will become explicit below.

\subsection{Hamiltonian on the links}
\label{sec:links}

On each of the links of the honeycomb lattice, labeled $\alpha$, we
place four spins that we collect into the 4-spinor $\sz_\alpha$:
\begin{align}
    \sz_\alpha = \begin{bmatrix}
    \sz_{\alpha,1}\\ \sz_{\alpha,2}\\ \sz_{\alpha,3}\\
    \sz_{\alpha,4}
  \end{bmatrix}
  \,.
\end{align}
In the $z$-basis, each $\sz_\alpha$ will form the basis
representation of the group elements as in Eq.(\ref{eq:vector_rep})
and each transforms under $r$ and $\ell$ as in in
Eq.(\ref{eq:rl_multiplication}). To ensure that we will be able to
project to the even-parity subspace we introduce four additional
spins, represented by the 4-spinor $\nz_\alpha$,
\begin{align}
    \nz_\alpha = \begin{bmatrix}
    \nz_{\alpha,1}\\ \nz_{\alpha,2}\\ \nz_{\alpha,3}\\
    \nz_{\alpha,4}
  \end{bmatrix}
  \,,
\end{align}
which are coupled to the $\sz_\alpha$ via the Ising interactions
\begin{align}
  H^{\rm Ising}_{\text{leg},\alpha}
  =
  -K\;{\nz_\alpha}^\top \; w \; \sz_\alpha\;,
  \label{eq:Ising-w}
\end{align}
with couplings proportional to the Hadamard matrix in
Eq.(\ref{eq:leg_w}). These couplings are the same as those described
in Ref.~\onlinecite{Z2-CGS}, and they favor the even-parity subspace
($\sz_{\alpha,1}\, \sz_{\alpha,2}\, \sz_{\alpha,3}\,
\sz_{\alpha,4}=1$). These couplings are invariant under monomial
transformations. For example if
\mbox{$\sz_\alpha\rightarrow r(g)\,\sz_\alpha$} then
\mbox{${\nz_\alpha}^\top \rightarrow {\nz_\alpha}^\top\,\ell(g)$}
preserves the Hamiltonian because of the automorphism of $w$ in
Eq.(\ref{eq:w-invariance}).

To the Ising couplings in Eq.~\eqref{eq:Ising-w} we can add uniform
transverse fields, which are invariant under the
transformations $r$ and $\ell$, because they are invariant under both flips of the
$z$-components and permutations of matter spins. The general Hamiltonian on each leg is then:
\begin{align}
  H_{\text{leg},\alpha}
  =&-K\,{\nz_\alpha}^\top \; w \; \sz_\alpha
  \nonumber\\
  &- \Gamma_\sigma \sum_i\sx_{\alpha,i} - \Gamma_\nu \sum_i\nx_{\alpha,i}
  \;.
  \label{eq:H_leg}     
\end{align}

\subsection{Hamiltonian on the vertex}
\label{sec:vertex}

In the second step of the construction we recursively collect the
three legs meeting at a vertex, introduce matter spins at the vertex,
and construct a gauge-matter coupling matrix $W$. $W$ will be
invariant under non-Abelian gauge transformations and will also favor
the configurations in which the product of the quaternions on those
three legs equals the identity.

We collect the three 4-spinors on each leg into a 12-spinor
$[\,\sz\,]$, 
\begin{align}
    [\,\sz\,] = \begin{bmatrix}
    \sz_1\\ \sz_2\\ \sz_3
    \end{bmatrix}\,.
\end{align}
We also define $64$-spinor of matter spins
\begin{align}
    [\,\mz\,] = \begin{bmatrix}
    \mz_{f_1, h_1}\\ \vdots \\ \mz_{f_{64}, h_{64}}
  \end{bmatrix}\,,
\end{align}
where the 64 states correspond to the choices of 64 triplets of group
elements that multiply to 1, parametrized as $f_i$, $h_i$,
and $(h_i\,f_i)^{-1}$, $i=1,\dots, 64$.

Now we couple the gauge and matter spins as follows
\begin{align}
  H^{\rm Ising}_{\text{vertex}}
  =
  -J\,[\,\mz\,]^\top \; W \; [\,\sz\,]
  \;.
  \label{eq:H_junction}
\end{align}
The the $64\times 12$ matrix $W$ is the interaction matrix defined by
\begin{align}
W
  =\frac{1}{4}\;
  \begin{bmatrix}
    v(f_1) & v(h_1) & v((f_1h_1)^{-1})\\[6pt]
     v(f_2) & v(h_2) & v((f_2h_2)^{-1})\\
    \vdots & \vdots & \vdots \\
     v(f_{64}) & v(h_{64}) & v((f_{64} h_{64})^{-1})\\
  \end{bmatrix}
  \;,
  \label{eq:W}
\end{align}
Each $v$ is exactly as defined by Eq.~(\ref{eq:vector_rep}) and
therefore the interactions are $\pm J$, i.e., ferromagnetic or
anti-ferromagnetic.  This interaction matrix enumerates all
configurations that satisfy the flux relation on the vertex in
Fig.~\ref{fig:QuantumDouble}. In each row we have that the product of
the quaternion group elements represented by the three $v$'s in each
row \textit{from left to right} is 1 (order matters).  The number of
rows exhausts the list of all possible leg configurations, matching
the number of matter spins.

We can also apply transverse and longitudinal fields to the matter
spins, while preserving the combinatorial symmetry, because they are
invariant under permutations of the matter spins at each vertex. The
more general Hamiltonian is then:
\begin{align}
  H_{\text{vertex}} =
  & -J\,[\,\mz\,]^\top \; W \; [\,\sz\,]
                         \nonumber\\
  &-\Gamma_\mu \sum_{i} \mx_{f_i,\,h_i}-H_\mu\sum_{i}\mz_{f_i,\,h_i}\;.
  \label{eq:H_junction}                
\end{align}

On the full lattice, the Hamiltonian is the sum of the leg terms in
Eq.~(\ref{eq:H_leg}) and vertex terms in Eq.~(\ref{eq:H_junction}). We
denote the legs emanating from each vertex $s$ by $\alpha(s)$ so that
the full Hamiltonian is:
\begin{align}
    H=\sum_s H_{\text{leg},\alpha(s)}+H_{\text{vertex,s}}\;.
  \label{eq:H_lattice} 
\end{align}

\subsection{Non-Abelian combinatorial gauge symmetry}
\label{sec:sym}

\begin{figure}[bh]
\centering
\includegraphics[width=.45\textwidth]{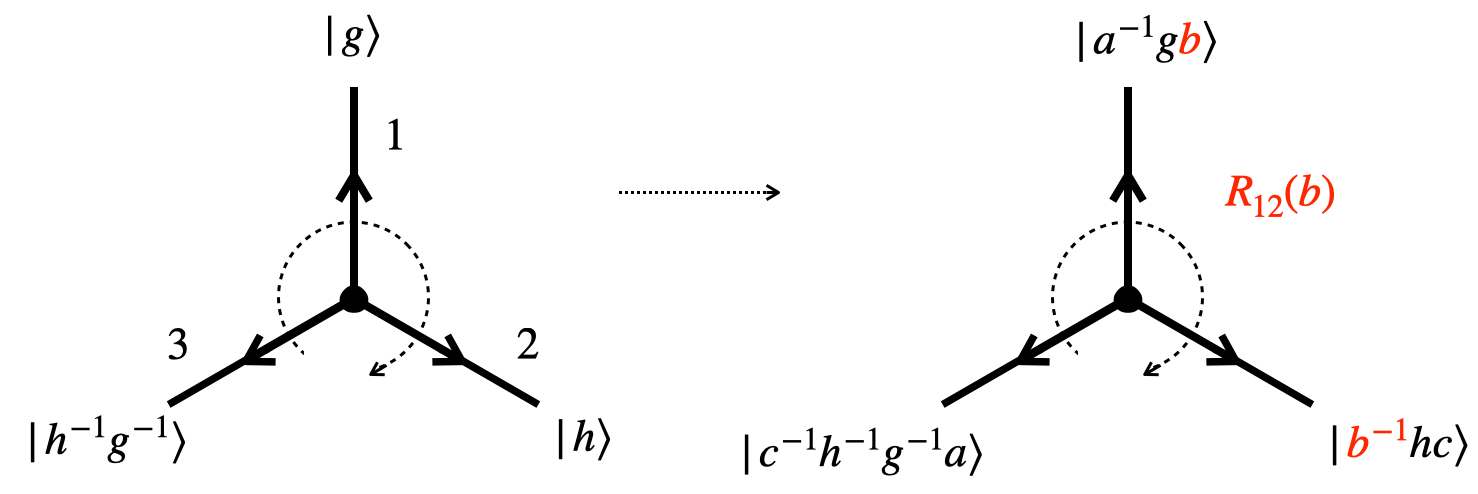}
\caption{Most general transformation of the hexagonal vertex that preserves the flux, where $a$, $b$ and $c$ are group elements. $R_{12}(b)$ is highlighted in red as an example.} 
\label{fig:VertexTransformation} 
\end{figure}

Now we show that each term in the full lattice Hamiltonian Eq.~\eqref{eq:H_lattice} is invariant under
local non-Abelian transformations. Let us start with the $ZZ$-term of
the vertex Hamiltonian Eq.~\eqref{eq:H_junction}, and consider
transformations as depicted in
Fig.~\ref{fig:VertexTransformation}. Consider three $12\times 12$ block
diagonal matrices, acting on the 12-spinor $[\,\sz\,]$, for group
elements $a$, $b$, and $c$:
\begin{subequations}
\label{eq:RRR}
\begin{align}
  R_{31}(a)=&
  \begin{bmatrix}
    \ell(a^{-1}) &0&0\\
    0&\openone_4&0\\
    0&0&r(a)
  \end{bmatrix}
  \\
  \nonumber\\
    R_{12}(b)=&
  \begin{bmatrix}
    r(b)&0&0\\
    0&\ell(b^{-1})&0\\
    0&0&\openone_4
  \end{bmatrix}
  \\
  \nonumber\\
  R_{23}(c)=&
  \begin{bmatrix}
    \openone_4 &0&0\\
    0&r(c)&0\\
    0&0&\ell(c^{-1})
  \end{bmatrix}
         \;,
\end{align}
\end{subequations}
where $\openone_4$ is the $4\times 4$ identity matrix and the matrices
$\ell(h)$ and $r(h)$ are the $4\times4$ left/right representations of
quaternions. The matrices in Eq.~\eqref{eq:RRR} act on two of the
three legs on the vertex, one by multiplying by a group element on the
left, and one by the (inverse) group element on the right. Each of
these transformations preserve the zero flux condition on the vertex
(i.e., that the three group elements multiply to the identity). Each
$R_{ij}$ obeys the quaternion algebra $R_{ij}(g)R_{ij}(h)=R_{ij}(gh)$
for all $i,j$, which follows from $r(g)r(h)=r(gh)$ and
$\ell(g)\ell(h)=\ell(hg)$. When one of the three legs is different,
the $R$'s commute, which follows from the commutation of the $r$ and
$\ell$ block matrices. The latter property ensures that we will be
able to insert charges locally because the two operations of inserting a group element into one hexagon and another group element into a neighboring hexagon in Fig.\ref{fig:Lattice} will commute, as they correspond to multiplication on the left and right.

For example, the transformation $R_{12}(b)$ is pictured in
Fig.~\ref{fig:VertexTransformation}b. Generically $R_{12}(b)$ has the
effect of mapping group elements $f_i$ and $h_i$ by
\begin{align}
  f_i &\rightarrow f_i\,b
  \nonumber\\
  h_i &\rightarrow b^{-1}\,h_i
\end{align}
for all lines $i=1,\dots, 64$ of the matrix $W$ in
Eq.~\eqref{eq:W}. Explicitly:
\begin{align}
    W\,R_{12}(b) &= 
    \begin{bmatrix}
    v(f_1)\;r(b) & v(h_1)\;\ell(b^{-1}) & v((f_1h_1)^{-1})\\[6pt]
     v(f_2)\;r(b) & v(h_2)\;\ell(b^{-1}) & v((f_2h_2)^{-1})\\
    \vdots & \vdots & \vdots \\
     v(f_{64})\;r(b) & v(h_{64})\;\ell(b^{-1}) & v((f_{64} h_{64})^{-1})\\
  \end{bmatrix}
  \nonumber\\\nonumber\\
  &\equiv
    \begin{bmatrix}
    v(f_1b) & v(b^{-1}h_1) & v((f_1h_1)^{-1})\\[6pt]
     v(f_2b) & v(b^{-1}h_2) & v((f_2h_2)^{-1})\\
    \vdots & \vdots & \vdots \\
     v(f_Nb) & v(b^{-1}h_N) & v((f_N h_N)^{-1})\\
  \end{bmatrix}
  \;,
  \\
  \nonumber
\end{align}
where we used Eq.(\ref{eq:rl_multiplication}). Note that
$W\,R_{12}(b)$ contains all the same 64 lines of $W$, but permuted,
i.e., we can write:
\begin{align*}
    WR_{12}(b) = L_{12}(b)\,W\;,
\end{align*}
where $L_{12}(b)$ is a $64\times 64$ permutation matrix. In general,
for any matrix $R$ that is a product of the $R_{ij}$ in Eq.(\ref{eq:RRR}) there is a corresponding
permutation matrix $L$ such that
\begin{align}
    L^\top\,W\,R = W\;,
\label{eq:vertex_Hadamard}
\end{align}
where we used the fact that the inverse of any permutation matrix is
its transpose. The automorphism in Eq.~(\ref{eq:vertex_Hadamard})
implies that the first term (first line) in the Hamiltonian
Eq.~\eqref{eq:H_junction} is invariant under the local non-Abelian
gauge transformation. The second and third terms (second line) of
Eq.~\eqref{eq:H_junction} are invariant under the permutation of the
64 matter spins $\mu$, given that the couplings $\Gamma_\mu$ and
$H_\mu$ are uniform.

\textit{Consistency of the leg and vertex Hamiltonians}: It is important to point out that the Hamiltonians as constructed in Eqs.~(\ref{eq:H_leg}) and (\ref{eq:H_junction}) are internally consistent. We have used two automorphisms -- one on the legs in Eq.~(\ref{eq:gauge}) and one on the vertex in Eq.~(\ref{eq:vertex_Hadamard}). In fact they are consistent by construction because the vertex transformations $R$ in Eq.~(\ref{eq:RRR}) are themselves composed of $r$ and $\ell$ matrices acting in the legs.     

The transverse field terms in the Hamiltonian are not spoiled by these transformations because the monomial matrices are implemented by rotations around the $x$-axis and permutations. The fields $\Gamma_\sigma$ and $\Gamma_\nu$ are uniform and hence invariant under permutations of gauge or matter spins, respectively, on each leg (permuting spins \textit{across} legs would not be allowed as it destroys the geometry of the lattice).

\section{The ground state manifold}

Here we establish that the ground state of the Hamiltonian
Eq.~\eqref{eq:H_lattice} is an equal amplitude superposition of all the
states satisfying the zero flux condition that are accessible from a
reference configuration by local plaquette operations. (Ground states
in distinct topological sectors are connected to different reference
states.) To arrive at this result, we first consider the spectrum of a
single vertex with the transverse fields switched off in Hamiltonian
Eq.~\eqref{eq:H_junction}, and show in Appendix~\ref{sec:details} that
the lowest energy manifold of states is comprised by those respecting
the zero flux condition.

To begin with, we need to ensure that even parity on each leg holds such that the quaternion representation in Eq.~(\ref{eq:vector_rep}) is valid. This condition is $K>5J/2$ applied to the couplings between gauge and matter spins in the Hamiltonian~(\ref{eq:H_lattice}). We will also require that the longitudinal field $H_\mu>0$ in order to favor the unit flux state on each vertex. Both conditions are derived in Appendix~\ref{sec:details}. 

Upon turning on the transverse fields, we obtain transition matrix
elements between the states satisfying the zero-flux condition in all
vertices of the lattice. The lowest order terms generated in the
perturbative expansion are hexagonal plaquette operators, $A_g(p)$,
which multiply the six links visited by the small loop by a sequence
of alternating elements $g$ and $g^{-1}$ of $\Q8$, as depicted in
Fig.~\ref{fig:Lattice}. 

The amplitudes in front of each hexagonal plaquette operator $A_g$
depend on the group element $g$. The operator $A_1$ has a second order
in $\Gamma$'s contribution; the operator $A_{-1}$ has the highest
order in $\Gamma$'s coefficient; and all operators $A_g$, $g\ne 1,-1$,
enter with equal coefficients (and at equal orders in perturbation
theory). That this is the case can be seen by inspection of the basic
representation of the group elements in
Eq.~\eqref{eq:vector_rep}. Multiplication by $g=-1$ necessarily flips
all spins that represent any group element. On the other hand,
multiplication by any $g\neq 1,-1$ flips exactly two spins (and
similarly for $g^{-1}$). The energy of matter spins that are permuted
by the operation of inserting a flux $g$ follows the same pattern
because the ground state manifold is invariant under combinatorial
gauge symmetry. Therefore, the effective plaquette Hamiltonian can be
written as
\begin{align}
  H_{\rm plaquette} =
  -\sum_p \sum_{g\in \Q8} \beta_g\;A_g(p)
  \;,
\end{align}
where $p$ is a plaquette and
$0 < \beta_{-1} < \beta_{g\ne -1,1} < \beta_1 = 1$ (the explicit form
of $\beta_g$ is not required for this argument).  $H_{\rm plaquette}$
commutes with the flux condition, as the $A_g(p)$ are the generators
of the gauge symmetry.

In Kitaev's quantum double construction in Ref.~\onlinecite{Kitaev},
the weights $\beta_g$ entering the plaquette Hamiltonian are chosen to be
all equal. This equal-weight choice is convenient but not required:
the necessary condition is that the weights $\beta_g>0$, to guarantee that
the quantum ground state is an equal superposition of all zero-flux
states (within a given topological sector). The equal amplitude
superposition of such states is an eigenvector of each $A_g(p)$ and
hence of $H_{\rm plaquette}$; that this superposition is the lowest
energy (ground state) eigenvector of $H_{\rm plaquette}$ follows from
the Perron-Frobenius theorem if the weights $\beta_g$ are all positive.

We thus arrive at the quantum ground state of the non-Abelian quantum
double model corresponding to the quaternion group, using only 1- and
2-body interactions. In the next section we discuss a superconducting
circuit with the elements needed to realize the Hamiltonian
Eq.~\eqref{eq:H_lattice}. If one is considering a physical system, the
gap in the spin Hamiltonian would be prohibitively small given the
high order of the plaquette terms. However, there is a possibility
that an alternate construction may in fact have a larger (possibly
non-perturbative) gap as we discuss in the next section.

\section{Superconducting circuit realization}

Instead of spins we can contemplate an array of superconducting wires connected by Josephson junctions.  Such a construction would follow closely that described in
Ref.~\onlinecite{sc-waffle}, in which the 2-body Ising terms in the
equivalent of Eq.~\eqref{eq:Ising-w} were obtained using a ``waffle''
where 4 vertical superconducting wires encode 4 matter degrees of
freedom, and 4 horizontal wires encode 4 gauge degrees of
freedom. Josephson junctions couple the horizontal and
vertical wires at their 16 intersections. The $\pm$ signs that enter
the matrix $w$ in Eq.~\eqref{eq:leg_w} can in principle be realized, for
instance, with regular Josephson couplings ($+$) or with
$\pi$-junctions ($-$).

The larger, $W$ matrix of couplings entering in \eqref{eq:H_junction}
and enforcing the zero-flux condition at the junctions of the
honeycomb lattice would require a larger assembly, with a $64\times 12$
mesh. The 64 wires in the construction correspond to the rows of
Eq.~\eqref{eq:W}, while the other 12 perpendicular wires correspond to
the columns of the matrix; each element of the $64\times 12$ $W$
matrix contains either a $+$ or a $-$ entry, which dictates whether
the intersection of the vertical and horizontal wires (associated to
the column and row) requires a regular Josephson coupling or a
$\pi$-junction.

While the construction may appear complex, with a multitude of
superconducting wires with specific types of junctions at the
$64\times 12$ crossings, one may hope that the perceived complexity
may diminish if very large scale integration of superconducting
circuits could ever follow the steps of semiconductor device
integration. Nonetheless, complex as it may, the Hamiltonian of this
system realizes {\it exactly} the non-Abelian gauge symmetry of the
quaternion quantum double, with physical interactions.

In the limit where the superconducting wires are small,
and voltage biases are tuned so two nearly degenerate charge states
are favored in each wire, the system would become a non-Abelian
generalization of the WXY model introduced in
Ref.~\onlinecite{sc-waffle}. It is not a priori obvious whether this model
would be gapped or gapless; if gapped, the scale of the gap would be
non-perturbative on the only remaining energy scale in the problem,
the Josephson coupling.

\section{Conclusions}

In this paper we constructed a physical spin Hamiltonian, with 1- and
2-body spin-spin interactions, which realizes a non-Abelian quantum
double for the quaternion group, $\Q8$, on a honeycomb lattice. To
physically represent the quaternion states, we used 4 spin-1/2 degrees
of freedom. We separated the corresponding 16-dimensional Hilbert
space into two subspaces of even and odd parity, and then we used the
8 even parity states to represent the 8 quaternions. We introduced
additional matter spin degrees of freedom on the links and vertices of
the honeycomb lattice: (i) to select the even sectors for the gauge
degrees of freedom on each link; and (ii) to favor the state in which
the three quaternion variables on the links connected to a lattice
site multiply to the identity, i.e., to select the zero flux condition
at each site as the ground state. We showed that the Hamiltonian,
which includes 2-body terms that couple gauge and matter spins as well
as transverse and longitudinal 1-body terms, possess an {\it exact}
combinatorial gauge symmetry associated with the quaternion
group. Notably, the spin model that we write has ferromagnetic and
anti-ferromagnetic $ZZ$ interactions, plus longitudinal and transverse
fields, and is thus an explicit example of a transverse Ising-type
model that {\it does not} have a {\it sign} problem and yet realizes a
non-Abelian topological phase. (The sign problem is
  absent for simulations working on the $Z$ basis, in which the $ZZ$
  terms and logitudinal fields are diagonal, with the transverse
  fields determining the off-diagonal transition matrix elements
  between different configurations, always of the same sign,
  independent of the initial or final configuration.) We observe, however, that although the system is
  sign-problem-free, numerical simulation would still be difficult, as
  the number of spins per unit cell is large.

Our Hamiltonian would not be expected in naturally occurring
materials. Instead, a physical realization of it would require a
programmable quantum device. While it requires a large number of
spin-1/2 degrees of freedom per unit cell to realize the system with
the exact non-Abelian combinatorial gauge symmetry, the fact that it
requires only 1- and 2-body spin-spin interactions makes it realistic
to expect that it could be in fact programmed in a device that has the
required number of qubits and the proper connectivity among them.

We also discussed a superconducting quantum circuit realization with
the same non-Abelian combinatorial gauge symmetry. This type of
circuit, even if it requires many elements, could conceivably be realized if
large-scale integration of superconducting quantum circuits continues
to advance.
  

\begin{acknowledgments}

  This work was supported by DOE Grant No. DE-FG02-06ER46316.
  
\end{acknowledgments}


\appendix

\section{Spectrum of a single vertex without transverse fields}
\label{sec:details}

Here we consider the spectrum of a single vertex with the transverse
fields switched off in Hamiltonian Eq.~\eqref{eq:H_junction}, and show
that that the lowest energy manifold of states is comprised by those
respecting the zero flux condition.

\subsection{Even-parity spin states}
\label{sec:even-parity}

Without loss of generality, we focus on the case of a vertex in
sublattice A, where the three group elements $g_1, g_2$ and $g_3$
defined on the edges of the vertex multiply {\it clockwise}. Let $S_+$
be the set of all triplets of group elements in $\Q8$,
$(g_1, g_2, g_3)$, such that $g_1\,g_2\,g_3=1$. $|S|=|\Q8|^2=8^2$, as
choosing $g_1$ and $g_2$ fixes $g_3$. Similarly, let $S_-$ be the set
of all triplets of group elements in $\Q8$, $(g_1, g_2, g_3)$, such
that $g_1\,g_2\,g_3=-1$. We define $S=S_+\cup S_-$, and $\bar S$ the
complement of $S$, i.e., all the triplets such that
$g_1\,g_2\,g_3 \ne \pm 1$.  ($|\bar S|=6\times 8^2$.)

By enumerating all the $8^3$ triplets $(g_1, g_2, g_3)$ and using
their associated spin representation $[\,\sz\,]$, we compute the
energies given by Eq.~\eqref{eq:H_junction} in the absence of the
transverse field. Let us first compute the contributions from the
2-spin interaction terms (proportional to $J$), namely
$E_J = -J\,[\,\mz\,]^\top \; W \; [\,\sz\,]$. We obtain for any
triplet in $S_+$ that
\begin{align}
  \label{eq:HS}
  E_{S_+} &=
  (-1J)\,\sum_{i=1}^{21} \mz_{\pi_i}
  +
  (0J)\,\sum_{i=22}^{45} \mz_{\pi_i}
  \nonumber\\
  &\;+
  (+1J)\,\sum_{i=46}^{63} \mz_{\pi_i}
  +
  (+3J)\,\sum_{i=64}^{64} \mz_{\pi_i}
  \;,
\end{align}
for $\pi$ a permutation of the 64 indices. That is, the spectrum is
independent of the state, upon permutations of the $\mu$ spins. The
minimum energy is $E_{s_+}^{\rm min}=-42J$, attained when 21 of the
$\mz$'s are positive, 19 are negative, and 24 can be either positive
or negative (this degeneracy can be lifted with a field, see below).

Similarly, for triplets in $S_-$, the Hamiltonian reduces to
\begin{align}
  \label{eq:HS}
  E_{S_-} &=
  (+1J)\,\sum_{i=1}^{21} \mz_{\pi_i}
  +
  (0J)\,\sum_{i=22}^{45} \mz_{\pi_i}
  \nonumber\\
  &\;+
  (-1J)\,\sum_{i=46}^{63} \mz_{\pi_i}
  +
  (-3J)\,\sum_{i=64}^{64} \mz_{\pi_i}
  \;.
\end{align}
Notice that the coefficients are opposite to those in the case of
triplets in $S_+$. But the minimum energy remains at
$E_{s_-}^{\rm min}=-42J$, attained when 21 of the $\mz$'s are negative,
19 are positive, and 24 can be either positive or negative.

Finally, for triplets in $\bar S$, the Hamiltonian reduces to
\begin{align}
  \label{eq:HSbar}
  E_{\bar S} &=
  (-2J)\,\sum_{i=1}^{3} \mz_{\pi_i}
  +
  (-1J)\,\sum_{i=4}^{15} \mz_{\pi_i}
  +
  (0J)\,\sum_{i=16}^{49} \mz_{\pi_i}
  \nonumber\\
  &\;+
  (+1J)\,\sum_{i=50}^{61} \mz_{\pi_i}
  +
  (+2J)\,\sum_{i=62}^{64} \mz_{\pi_i}
\end{align}
with minimum energy is $E_{\bar s}^{\rm min}=-36J$.

We thus separate the manifold of states into two classes, those in
$S_+$ and $S_-$ and those in $\bar S$ by a value $6J$. The minima for
$S_+$ and $S_-$ states can be split by applying a uniform longitudinal
field to the matter spins. A positive field $H_\mu$, as included in
Eq.~\eqref{eq:H_junction}, lowers the energy of the $S_+$ manifold
with respect to the $S_-$ one by a value $4\,H_\mu$
[$=(21-19)H_\mu+(19-21)H_\mu$].

There are degeneracies (which can be read from the number of spins
multiplying the coefficient $0J$ above). These degeneracies can be
lifted by the longitudinal field, or also by a transverse field.

\subsection{Inclusion of odd-parity spin states}

In the above we only included the even-parity states (in terms of four
spins) associated to the 8 elements of the quaternion
group. Explicitly, we considered the 8 states in
Eq.~\eqref{eq:vector_rep}. There are the remaining 8 odd-parity states
that are physical in terms of the four spins but do not correspond to
quaternions. These are obtained from the ``good'' (even) states by,
schematically,
\begin{align}
  {\tilde v}({\rm odd}) =
  v({\rm even})\;
  \begin{bmatrix}
    -&0&0&0\\
    0&+&0&0\\
    0&0&+&0\\
    0&0&0&+
  \end{bmatrix}
  \;.         
\end{align}
Explicitly, we have the odd-parity states
\begin{alignat}{3}
  \vtil{+1}&=
  \begin{bmatrix}
    -+++
  \end{bmatrix}
  \qquad
  \vtil{-1}&=
  \begin{bmatrix}
    +---
  \end{bmatrix}
  \nonumber\\
  \vtil{+i}&=
  \begin{bmatrix}
    --+-
  \end{bmatrix}
  \qquad
  \vtil{-i}&=
  \begin{bmatrix}
    ++-+
  \end{bmatrix}
  \nonumber\\
  \vtil{+j}&=
  \begin{bmatrix}
    -+--
  \end{bmatrix}
  \qquad
  \vtil{-j}&=
  \begin{bmatrix}
    +-++
  \end{bmatrix}
  \nonumber\\
  \vtil{+k}&=
  \begin{bmatrix}
    +++-
  \end{bmatrix}
  \qquad
  \vtil{-k}&=
  \begin{bmatrix}
    ---+
  \end{bmatrix}
  &\;.
  \label{eq:vector_rep_tilde}
\end{alignat}

We obtain the energies of all possible states of a vertex, including
legs with both positive and negative parity, by exhaustively
enumerating all configurations and including the contributions from
Eqs.~\eqref{eq:H_leg}~and~\eqref{eq:H_junction} (with transverse
fields off). We obtain the following energies depending on the parity
of the three legs:
\begin{enumerate}[(i)]

\item $E=-42J - 12K$; parities on legs: $(1, 1, 1)$; 128 states (64 with
  $g_1g_2g_3=+1$ and 64 with $g_1g_2g_3=-1$).

\item $E=-36J -12K$; parities on legs: $(1, 1, 1)$; 384 states.

\item $E= -48J -6K$; parities on legs: $(-1, -1, -1)$; 512 states.

\item $E=-47J -10K$; parities on legs: $(-1, 1, 1)$, $(1, -1, 1)$,
  $(1, 1, -1)$; 512 states each.

\item $E=-40J -8K$; parities on legs: $(-1, -1, 1)$, $(-1, 1, -1)$,
  $(1, -1, -1)$; 512 states each.

\end{enumerate}

The set of states (i), with only even-parity links, form the lowest
energy manifold if the coupling constant $K$ is chosen such that
$K > 5J/2$. This result, together with those in
Sec.~\ref{sec:even-parity} above, establish that the manifold of
states corresponding to link elements $g_1, g_2,g_3\in \Q8$ with zero
flux (i.e., $g_1 g_2 g_3=1$) are separated from the other states by a
gap $4\,H_\mu$.

\bibliography{reference}

\end{document}